\documentclass{jps-cp}
\usepackage{txfonts} 

\usepackage{bm}
\def\mev{\text{ MeV}}
\def\gev{\text{ GeV}}

\title{Further Theoretical Analysis on the $K^{-} {}^{3} \text{He} \to \Lambda
  p n$ Reaction for the $\bar{K} N N$ Bound-State Search in the J-PARC
  E15 Experiment}

\author{Takayasu \textsc{Sekihara}$^{1}$, Eulogio \textsc{Oset}$^{2}$,
and Angels \textsc{Ramos}$^{3}$}

\inst{$^{1}$Advanced Science Research Center, Japan Atomic Energy Agency,
  Shirakata, Tokai, Ibaraki, 319-1195, Japan \\
  $^{2}$Departamento de F\'{\i}sica Te\'orica and IFIC, Centro
  Mixto Universidad de Valencia-CSIC, Institutos de Investigaci\'on
  de Paterna, Aptdo. 22085, 46071 Valencia, Spain \\
  $^{3}$Departament de F\'{\i}sica Qu\`antica i Astrof\'{\i}sica
  and Institut de Ci\`encies del Cosmos, Universitat de Barcelona,
  Mart\'i i Franqu\`es 1, 08028 Barcelona, Spain}

\email{sekihara@post.j-parc.jp}

\recdate{\today}

\abst{Based on the scenario that a $\bar{K} N N$ bound state is
  generated and it eventually decays into $\Lambda p$, we calculate
  the cross section of the $K^{-} {}^{3} \text{He} \to \Lambda p n$
  reaction, which was recently measured in the J-PARC E15 experiment.
  We find that the behavior of the calculated differential cross
  section $d ^{2} \sigma / d M_{\Lambda p} d q_{\Lambda p}$, where
  $M_{\Lambda p}$ and $q_{\Lambda p}$ are the $\Lambda p$ invariant
  mass and momentum transfer in the $(K^{-} , \, n)$ reaction in the
  laboratory frame, respectively, is consistent with the experiment.
  Furthermore, we can reproduce almost quantitatively the experimental
  data of the $\Lambda p$ invariant mass spectrum in the momentum
  transfer window $350 \mev /c < q_{\Lambda p} < 650 \mev /c$.  These
  facts strongly suggest that the $\bar{K} N N$ bound state was indeed
  generated in the J-PARC E15 experiment.}

\kword{$\bar{K} N N$ bound state, reaction calculation for the
  J-PARC E15 experiment}

\begin{document}
\maketitle

\section{Introduction}

Because the chiral $\bar{K} N$ interaction is strongly attractive and
dynamically generates the $\Lambda (1405)$~\cite{Kaiser:1995eg,
  Oset:1997it, Oller:2000fj, Oset:2001cn, Lutz:2001yb, Jido:2003cb},
it is natural to extend the idea from the $\bar{K} N$ bound state
[$\Lambda (1405)$] to the $\bar{K} N N$ bound state.  The $\bar{K} N
N$ bound state is a good ground to apply techniques of few-body
calculations and to investigate further properties of the $\bar{K} N$
interaction.  Until now a lot of effort has been put into theoretical
predictions and experimental searches for the $\bar{K} N N$ bound
state (see Ref.~\cite{Nagae:2016cbm} for the status up to 2015).

Recently, in the J-PARC E15 experiment the cross section of the $K^{-}
{}^{3} \text{He} \to \Lambda p n$ reaction was measured and a peak
structure was found in the $\Lambda p$ invariant mass spectrum around
the $K^{-} p p$ threshold~\cite{Sada:2016nkb, Ajimura:2018iyx}.  One
of the biggest advantages of this reaction is to put $\bar{K}$
directly into the nucleus to generate the $\bar{K} N N$ bound state.
Thanks to that, we can theoretically trace the behavior of the
$\bar{K}$ in the $K^{-} {}^{3} \text{He} \to \Lambda p n$
reaction~\cite{Sekihara:2016vyd}.  As a result, the first run data of
the J-PARC E15 experiment~\cite{Sada:2016nkb} were well reproduced in
the scenario that the $\bar{K} N N$ bound state is generated.

In this manuscript we perform further theoretical analysis on the
$K^{-} {}^{3} \text{He} \to \Lambda p n$ reaction measured in the
J-PARC E15 experiment, in particular focusing on its second run
data~\cite{Ajimura:2018iyx}.

\section{Formulation}

We calculate the cross section of the $K^{-} {}^{3} \text{He} \to
\Lambda p n$ reaction in the same manner as in our previous
paper~\cite{Sekihara:2016vyd} except for one thing: the treatment of
the $\bar{K} N \to \bar{K} N$ scattering amplitude at the first
collision denoted by $T_{1}$.  In Ref.~\cite{Sekihara:2016vyd} we
fixed $T_{1}$ as a real number by using the experimental values of the
$\bar{K} N \to \bar{K} N$ differential cross sections only at an
initial kaon momentum $1.0 \gev /c$ in the laboratory frame.  Namely,
we neglected the Fermi motion of the nucleons and fixed the invariant
mass of the first-collision $\bar{K} N$, $w_{1}$, to a unique value.
In contrast, now we take into account the Fermi motion of the nucleons
to evaluate $w_{1}$ and treat $T_{1}$ in a $2 \times 2$ matrix form in
terms of the partial waves:
\begin{align}
  T_{1} ( w_{1} , \, \bm{p}_{\rm out} , \, \bm{p}_{\rm in} )
  = g ( w_{1} , \, p_{\rm out} , \, p_{\rm in} , \, x )
  - i h ( w_{1} , \, p_{\rm out} , \, p_{\rm in} , \, x )
  \frac{( \bm{p}_{\rm out} \times \bm{p}_{\rm in} ) \cdot \bm{\sigma}}
       {p_{\rm out} p_{\rm in}} ,
\end{align}
where $\bm{p}_{\rm out}$ and $\bm{p}_{\rm in}$ are outgoing and
incoming momenta of $\bar{K}$ in the $\bar{K} N$ rest frame, $p_{\rm
  out,in} \equiv | \bm{p}_{\rm out,in} |$, $x \equiv \bm{p}_{\rm out}
\cdot \bm{p}_{\rm in} / ( p_{\rm out} p_{\rm in} )$, $\bm{\sigma}$ are
the Pauli matrices acting on the baryon spinors, and $g$ and $h$ are
expressed in terms of the partial-wave amplitudes $T_{L \pm}$ as
\begin{align}
  & g ( w , \, p_{\rm out} , \, p_{\rm in} , \, x )
  = \sum _{L = 0}^{\infty}
  \left [ ( L + 1 ) T_{L +} ( w , \, p_{\rm out} , \, p_{\rm in} )
    + L T_{L -} ( w , \, p_{\rm out} , \, p_{\rm in} )
    \right ]
  P_{L} ( x ) , 
  \\
  & h ( w , \, p_{\rm out} , \, p_{\rm in} , \, x )
  = \sum _{L = 1}^{\infty}
  \left [ T_{L +} ( w , \, p_{\rm out} , \, p_{\rm in} )
    - T_{L -} ( w  , \, p_{\rm out} , \, p_{\rm in} ) \right ]
  P_{L}^{\prime} ( x ) , 
\end{align}
with the Legendre polynomials $P_{L} ( x )$, $P_{L}^{\prime} ( x )
\equiv d P_{L} / d x$, and orbital angular momentum $L$.

Because the $\bar{K} N \to \bar{K} N$ scattering at the first
collision takes place with bound nucleons, it is better to treat the
partial-wave amplitudes $T_{L \pm}$ as functions of three independent
variables $w_{1}$, $p_{\rm out}$, and $p_{\rm in}$, i.e., as off-shell
amplitudes.  We here assume that the partial-wave amplitudes depend on
the momenta minimally required by the kinematics, i.e., they are
proportional to $( p_{\rm out} p_{\rm in} )^{L}$.  Under this assumption,
the partial-wave amplitudes can be easily evaluated from the on-shell
ones in the formula
\begin{align}
  T_{L \pm} ( w , \, p_{\rm out} , \, p_{\rm in} )
  = T_{L \pm}^{\text{on-shell}} ( w )
  \frac{( p_{\rm out} p_{\rm in} )^{L}}{[ p^{\text{on-shell}} ( w ) ]^{2 L}} ,
  \label{eq:min_on-shell}
\end{align}
where $p^{\text{on-shell}} ( w )$ is the on-shell momentum for the
$\bar{K} N$ system with the center-of-mass energy $w$.  In the present
study, we utilize the on-shell $\bar{K} N$ amplitudes $T_{L
  \pm}^{\text{on-shell}} ( w )$ of Ref.~\cite{Kamano:2014zba}, where
the authors calculated $\bar{K} N$ amplitudes up to the $F$ wave ($L =
4$) based on a dynamical coupled-channels model with phenomenological
SU(3) Lagrangians, to evaluate the amplitudes $T_{L \pm} ( w , \,
p_{\rm out} , \, p_{\rm in} )$.

\section{Numerical results}

\begin{figure}[t]
  \centering
  \includegraphics[width=7.6cm]{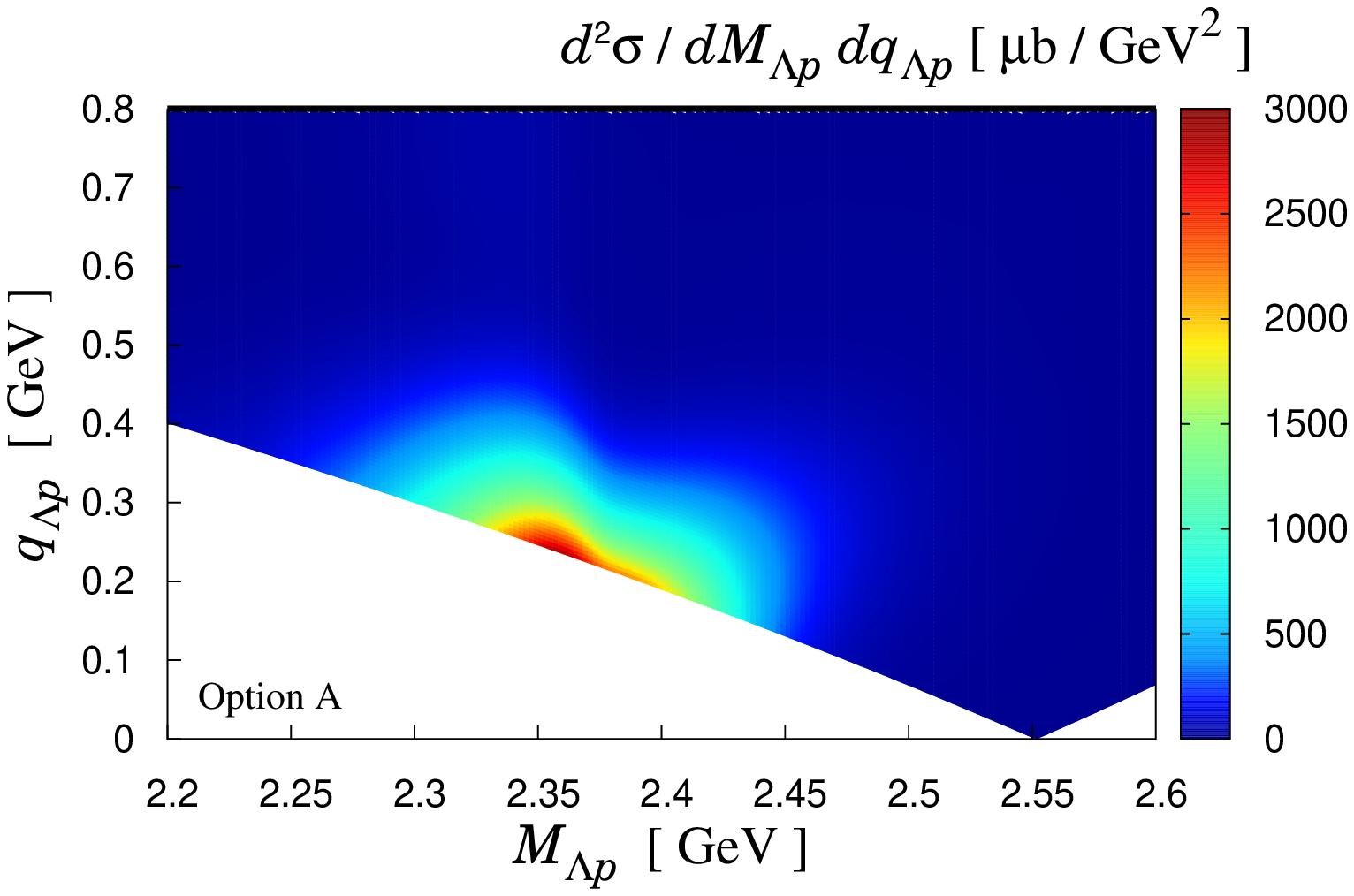} ~
  \includegraphics[width=7.6cm]{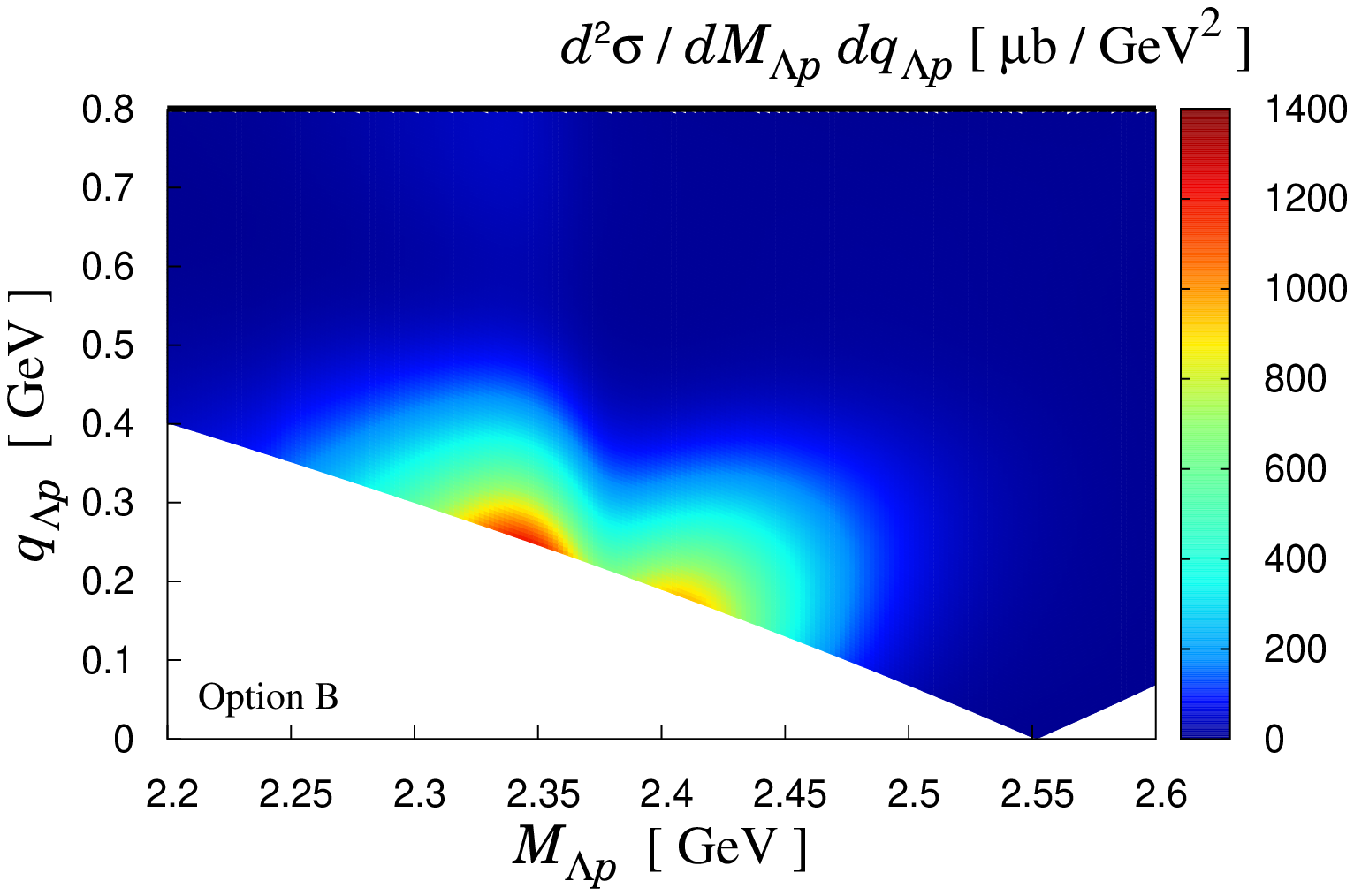}
  \caption{Differential cross section for the $K^{-} {}^{3} \text{He}
    \to \Lambda p n$ reaction in option A (left) and B (right).}
\label{fig:1}
\end{figure}

We now calculate the differential cross section $d^{2} \sigma / d
M_{\Lambda p} d \cos \theta _{n}^{\rm cm}$ of the $K^{-} {}^{3}
\text{He} \to \Lambda p n$ reaction, where $M_{\Lambda p}$ is the
$\Lambda p$ invariant mass and $\theta _{n}^{\rm cm}$ is the neutron
scattering angle in the global center-of-mass frame, in the scenario
that a $\bar{K} N N$ bound state is formed, decaying eventually into a
$\Lambda p$ pair.  The formulation is the same as in the previous
calculation~\cite{Sekihara:2016vyd} except for the $\bar{K} N \to
\bar{K} N$ amplitude at the first collision, which now takes the Fermi
motion into account.  We then multiply $d^{2} \sigma / d M_{\Lambda p}
d \cos \theta _{n}^{\rm cm}$ by $| \partial \cos \theta _{n}^{\rm cm}
/ \partial q_{\Lambda p} |$, where $q_{\Lambda p}$ is the momentum
transfer in the $(K^{-} , \, n)$ reaction in the laboratory frame, to
obtain the differential cross section $d^{2} \sigma / d M_{\Lambda p}
d q_{\Lambda p}$.  Note that, for a fixed $M_{\Lambda p}$, the
momentum transfer reaches its minimum at $\cos \theta _{n}^{\rm cm} =
1$ and increases as $\cos \theta _{n}^{\rm cm}$ decreases.

In Fig.~\ref{fig:1} we show the numerical result of the differential
cross section $d^{2} \sigma / d M_{\Lambda p} d q_{\Lambda p}$ in the
$\bar{K} N N$ bound-state scenario.  Here we take two options A and B
for the evaluation of the energy carried by the $\bar{K}$ (for the
details see Ref.~\cite{Sekihara:2016vyd}).  From Fig.~\ref{fig:1},
in both options A and B, we can see that the structure near the $K^{-}
p p$ threshold ($2.37 \gev$) is generated dominantly in the lower
$q_{\Lambda p}$ region, which corresponds to the condition of forward
neutron scattering.  In addition, we observe two trends in $d^{2}
\sigma / d M_{\Lambda p} d q_{\Lambda p}$; one goes from $M_{\Lambda
  p} = 2.35 \gev$ at $q_{\Lambda p} = 0.25 \gev$ to the upward
direction, and the other goes from the $K^{-} p p$ threshold at
$q_{\Lambda p} = 0.2 \gev$ to the upper-right direction.  As discussed
in Ref.~\cite{Sekihara:2016vyd}, the former is the signal of the
$\bar{K} N N$ bound state, and the latter is the contribution from the
quasi-elastic scattering of the $\bar{K}$ at the first collision.
Interestingly, the behavior of the two trends in $d^{2} \sigma / d
M_{\Lambda p} d q_{\Lambda p}$ was indeed observed in the second run
data of the J-PARC E15 experiment~\cite{Ajimura:2018iyx}, and hence
our result is consistent with the experimental result.

\begin{figure}[b]
  \centering
  \begin{minipage}{0.55\hsize}
  \includegraphics[width=7.6cm]{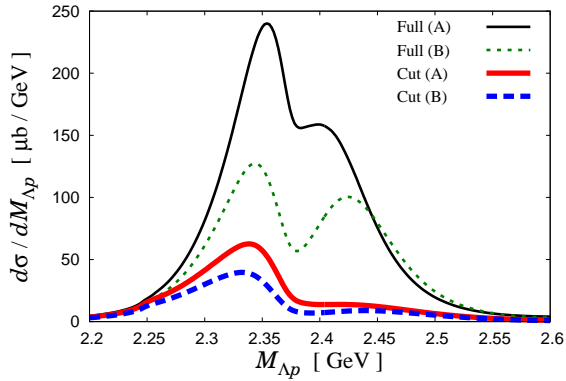}
  \end{minipage}
  \begin{minipage}{0.44\hsize}
    \caption{Invariant mass spectrum $d \sigma / d M_{\Lambda p}$ for
      the $K^{-} {}^{3} \text{He} \to \Lambda p n$ reaction in options
      A and B.  In the ``Cut'' case we restrict the momentum transfer
      to the region $350 \mev / c < q_{\Lambda p} < 650 \mev /c$.}
    \label{fig:2}
  \end{minipage}
\end{figure}

Next we integrate the differential cross section $d^{2} \sigma / d
M_{\Lambda p} d q_{\Lambda p}$ to obtain the invariant mass
spectrum $d \sigma / d M_{\Lambda p}$, which is shown in
Fig.~\ref{fig:2}.  When we take into account the whole region of the
momentum transfer, we obtain lines ``Full'' in Fig.~\ref{fig:2}.  In
this case, we observe a two-peak structure around the $K^{-} p p$
threshold as we found in Ref.~\cite{Sekihara:2016vyd}; the peak below
(above) the $K^{-} p p$ threshold originates from the $\bar{K} N N$
bound state (quasi-elastic scattering of the $\bar{K}$ at the first
collision).  Then, we restrict the momentum transfer to the region
$350 \mev < q_{\Lambda p} < 650 \mev$, as done in the experimental
analysis~\cite{Ajimura:2018iyx}, where they aimed at reducing the
kinematic peak above the $K^{-} p p$ threshold corresponding to the
quasi-elastic $\bar{K}$ scattering.  Our mass spectrum with the
momentum-transfer cut is plotted as lines ``Cut'' in Fig.~\ref{fig:2}.
With this cut, only the peak for the signal of the $\bar{K} N N$ bound
state survives.  In this sense, our result supports the validity of
the experimental cut of the momentum transfer.

\begin{figure}[t]
  \centering
  \begin{minipage}{0.55\hsize}
    \includegraphics[width=7.6cm]{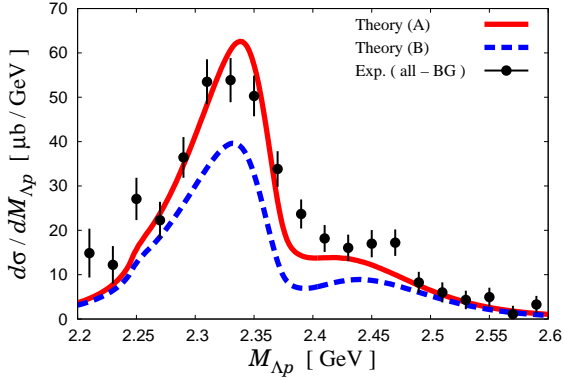}
  \end{minipage}
  \begin{minipage}{0.44\hsize}
    \caption{Comparison between theoretical and experimental results
      of the $\Lambda p$ invariant mass spectrum $d \sigma / d
      M_{\Lambda p}$ for the $K^{-} {}^{3} \text{He} \to \Lambda p n$
      reaction in the momentum transfer window $350 \mev / c <
      q_{\Lambda p} < 650 \mev /c$.  For the experimental data we
      subtract the background contribution in the experimental
      analysis~\cite{Ajimura:2018iyx}.}
    \label{fig:3}
  \end{minipage}
\end{figure}

Finally, we compare the calculated invariant mass spectrum $d \sigma /
d M_{\Lambda p}$ in the momentum transfer window $350 \mev <
q_{\Lambda p} < 650 \mev$ with the experimental one in the same
window.  The result is shown in Fig.~\ref{fig:3}.  As for the
experimental data, we subtract the background contribution in the
analysis of the experiment~\cite{Ajimura:2018iyx}, because we do not
take into account background but rather the generation of the $\bar{K}
N N$ bound state.  From Fig.~\ref{fig:3}, in both options A and B, we
can see that our calculation reproduces almost quantitatively the
experimental data of the $\Lambda p$ invariant mass spectrum with the
momentum transfer cut throughout a wide range of the $\Lambda p$
invariant mass.  The consistent behavior of the mass spectrum strongly
suggests that the $\bar{K} N N$ bound state was indeed generated in
the J-PARC E15 experiment.

\section{Summary}

In this manuscript we have investigated the origin of the peak
structure of the $\Lambda p$ invariant mass spectrum near the $K^{-} p
p$ threshold in the $K^{-} {}^{3} \text{He} \to \Lambda p n$ reaction,
which was recently observed in the J-PARC E15 experiment.  For this
purpose, we have calculated the cross section of the $K^{-} {}^{3} 
\text{He} \to \Lambda p n$ reaction and $\Lambda p$ invariant mass
spectrum based on the scenario that a $\bar{K} N N$ bound state is
generated and it eventually decays into $\Lambda p$.  As a result, we
have found that the behavior of the calculated differential cross
section $d ^{2} \sigma / d M_{\Lambda p} d q_{\Lambda p}$ is entirely
consistent with the experimental data.  In particular, the peak for
the quasi-elastic scattering of the $\bar{K}$ at the first collision
in the $\Lambda p$ invariant mass spectrum, which exists above the
$K^{-} p p$ threshold, is highly suppressed when we restrict the
momentum transfer to the region $350 \mev < q_{\Lambda p} < 650 \mev$,
as done in the experimental analysis~\cite{Ajimura:2018iyx}; with this
cut only the peak for the $\bar{K} N N$ bound state below the $K^{-} p
p$ threshold survives.  Furthermore, throughout a wide range of the
$\Lambda p$ invariant mass, our calculation reproduces almost
quantitatively the experimental mass spectrum with the momentum
transfer cut.  These findings strongly suggest that the $\bar{K} N N$
bound state was indeed generated in the J-PARC E15 experiment.


\begin{thebibliography}{99}

\bibitem{Kaiser:1995eg} 
  N.~Kaiser, P.~B.~Siegel and W.~Weise,
  Nucl.\ Phys.\ A {\bf 594}, 325 (1995).

\bibitem{Oset:1997it} 
  E.~Oset and A.~Ramos,
  Nucl.\ Phys.\ A {\bf 635}, 99 (1998).

\bibitem{Oller:2000fj} 
  J.~A.~Oller and U.~G.~Mei{\ss}ner,
  Phys.\ Lett.\ B {\bf 500}, 263 (2001).

\bibitem{Oset:2001cn} 
  E.~Oset, A.~Ramos and C.~Bennhold,
  Phys.\ Lett.\ B {\bf 527}, 99 (2002).

\bibitem{Lutz:2001yb} 
  M.~F.~M.~Lutz and E.~E.~Kolomeitsev,
  Nucl.\ Phys.\ A {\bf 700}, 193 (2002).

\bibitem{Jido:2003cb} 
  D.~Jido, J.~A.~Oller, E.~Oset, A.~Ramos and U.~G.~Mei{\ss}ner,
  Nucl.\ Phys.\ A {\bf 725}, 181 (2003).
  
  
\bibitem{Nagae:2016cbm} 
  T.~Nagae,
  Nucl.\ Phys.\ A {\bf 954}, 94 (2016).

\bibitem{Sada:2016nkb} 
  Y.~Sada {\it et al.} [J-PARC E15 Collaboration],
  Prog.\ Theor.\ Exp.\ Phys.\ {\bf 2016}, 051D01 (2016).

\bibitem{Ajimura:2018iyx} 
  S.~Ajimura {\it et al.} [J-PARC E15 Collaboration],
  Phys.\ Lett.\ B {\bf 789}, 620 (2019).

  
\bibitem{Sekihara:2016vyd} 
  T.~Sekihara, E.~Oset and A.~Ramos,
  Prog.\ Theor.\ Exp.\ Phys.\ {\bf 2016}, 123D03 (2016).
  
\bibitem{Kamano:2014zba} 
  H.~Kamano, S.~X.~Nakamura, T.-S.~H.~Lee and T.~Sato,
  Phys.\ Rev.\ C {\bf 90}, 065204 (2014).

\end{thebibliography}
\end{document}